\input{aipcheck}

\documentclass[
    ,final            
  ]
  {aipproc}

\layoutstyle{6x9}

\begin{document}

\title{Large $N_c$}


\author{Elizabeth E. Jenkins}{
  address={Department of Physics, University of California at San Diego, La Jolla, CA 92093-0319}
}

\begin{abstract}
The $1/N_c$ expansion of QCD with $N_c=3$ has been successful in explaining a wide variety of QCD phenomenology.  Here I focus on the contracted spin-flavor symmetry of baryons in the large-$N_c$ limit and
deviations from spin-flavor symmetry due to corrections suppressed by powers of $1/N_c$.  Baryon masses provide an important example
of the $1/N_c$ expansion, and successful predictions of masses of heavy-quark baryons continue to be tested by experiment.  The ground state charmed baryon masses have all been measured, and five of the eight ground state bottom baryon masses have been found.  Results of the $1/N_c$ expansion can aid in the discovery of the remaining bottom baryons.  The brand new measurement of the $\Omega_b^-$ mass by the CDF collaboration conflicts with the original $D0$ discovery value and is in excellent agreement with the prediction of the $1/N_c$ expansion.
\end{abstract}

\maketitle


\section{Introduction}

Many features of QCD can be understood by studying non-Abelian $SU(N_c)$ gauge theory in
an expansion in $1/N_c$ about the limit $N_c \rightarrow \infty$~\cite{thooft}.  The nonperturbative expansion
parameter $1/N_c$ is equal to $1/3$ for QCD.  In this talk, I concentrate on studying baryons in
the $1/N_c$ expansion.  

For large, finite $N_c$, a baryon is a color-singlet bound state of $N_c$
valence quarks completely antisymmetrized in the color indices of the quarks~\cite{witten}.  The baryon mass
for light quarks scales as $N_c \Lambda_{\rm QCD}$ and the size of the baryon is order $1/\Lambda_{\rm QCD}$,
independent of $N_c$.  The $N_c$-dependence of baryon-meson scattering amplitudes and couplings can be determined by studying quark-gluon diagrams.  The amplitude for a baryon and a meson to scatter to a baryon and $(n-1)$ mesons scales
as ${N_c}^{1 -n/2}$.  The baryon--antibaryon--$n$-meson vertex also scales as ${N_c}^{1 -n/2}$.  Notice that this scaling rule
implies that the baryon--antibaryon--1-meson vertex {\it grows} as $\sqrt{N_c}$.  Since the amplitude for baryon + meson $\rightarrow$
baryon + meson scattering is at most $O(1)$, the two leading pole diagrams (see Fig.~1) containing two 
baryon--antibaryon--1-meson vertices are
each $O(N_c)$, but the sum of the two diagrams is $O(1)$.  Thus, there must be an exact cancellation of the $O(N_c)$ contributions
of the two pole diagrams.  This cancellation requirement implies that there is a contracted spin-flavor symmetry for baryons
in the large-$N_c$ limit~\cite{dm}.  Additional consistency conditions lead to non-trivial constraints on $1/N_c$ corrections to the large-$N_c$ limit for baryons~\cite{dm,j,djm1}.

\begin{figure}
{\includegraphics[width=2truein]{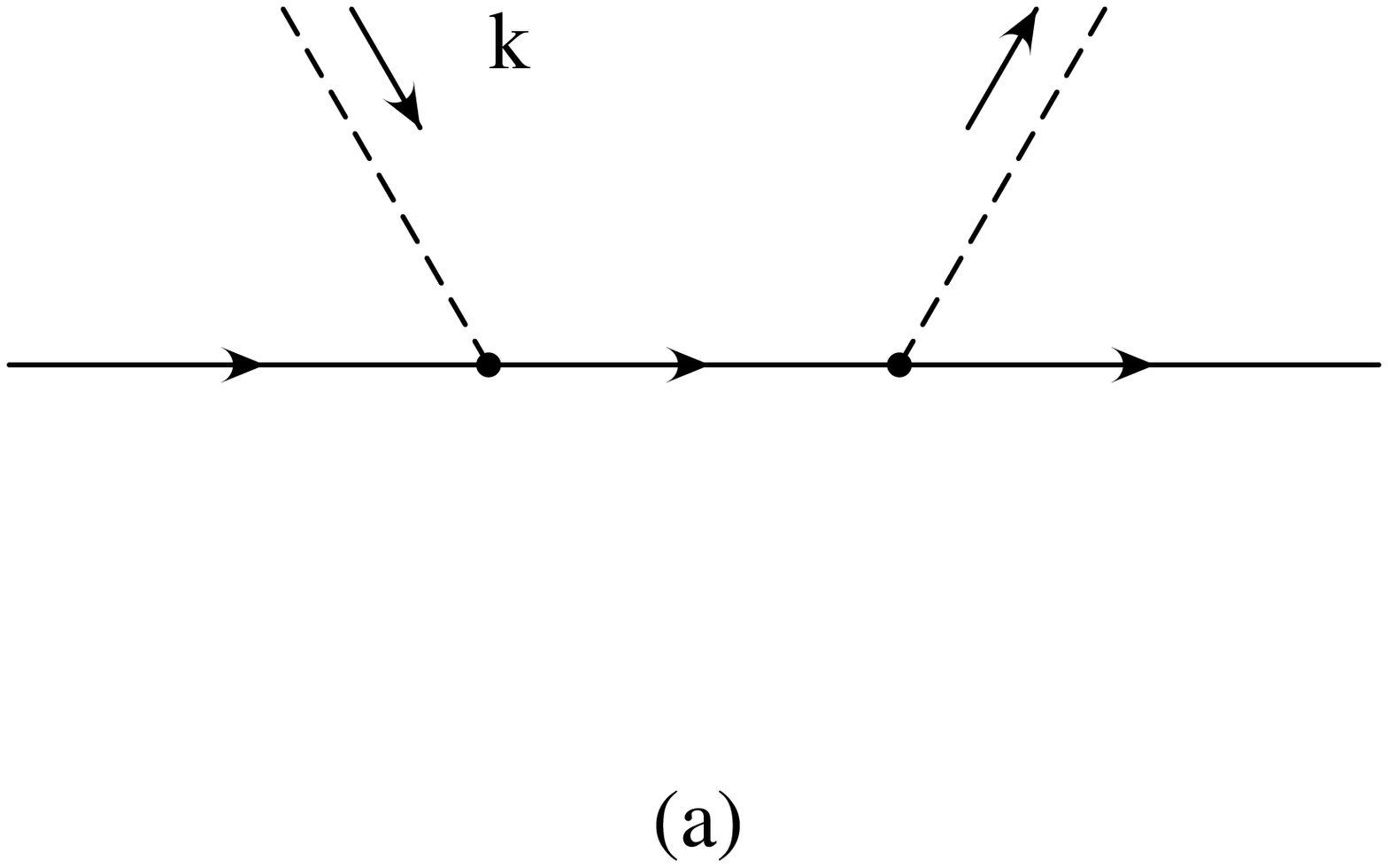}}
\phantom{space}
{\includegraphics[width=2truein]{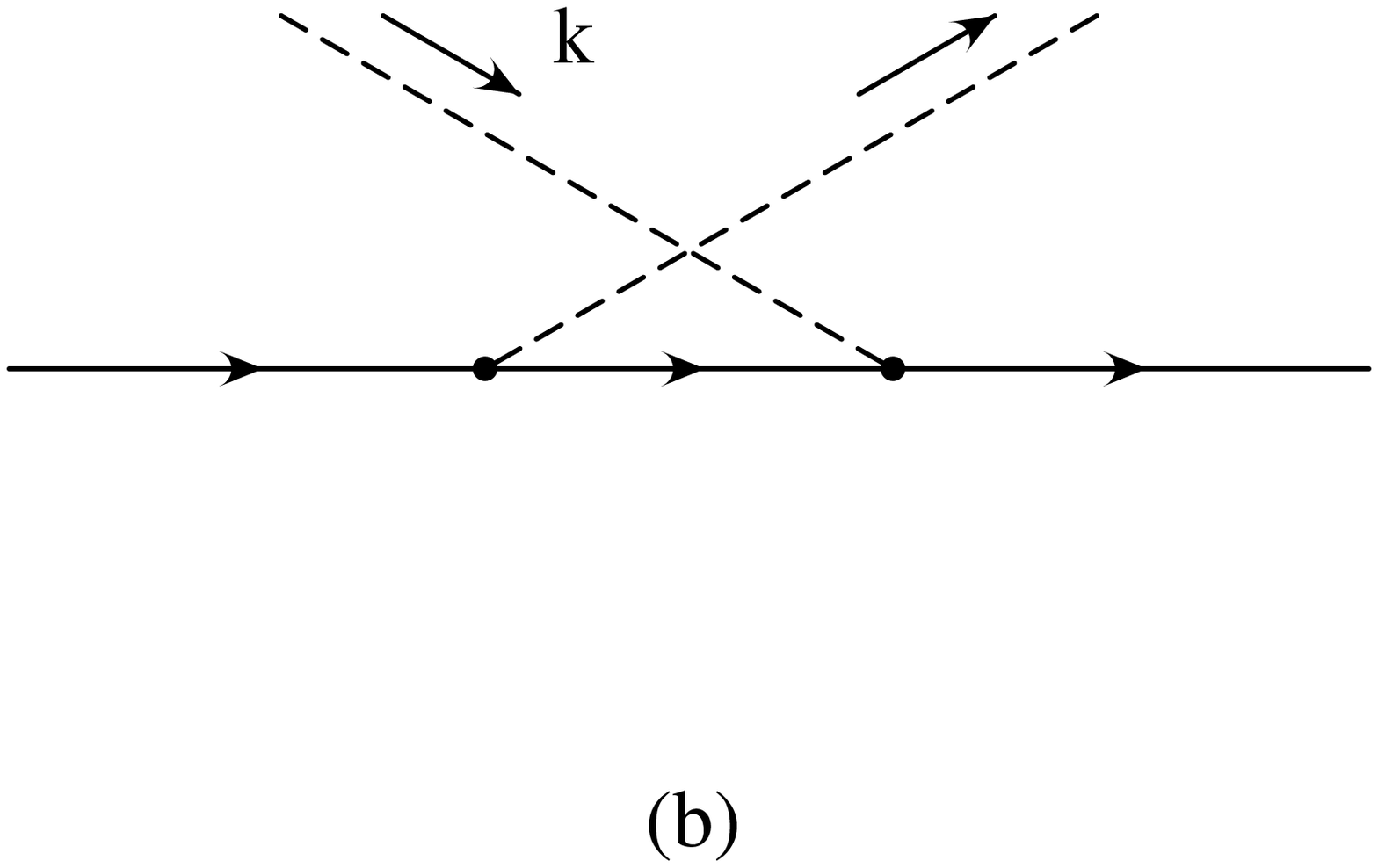}}
\caption{Two pole diagrams contributing to baryon + meson scattering.  Diagrams $(a)$ and $(b)$ each
contain two vertices of order $\sqrt{N_c}$, and contribute to the scattering amplitude at $O(N_c)$.  The
sum of the two diagrams, however, is $O(1)$ by the $N_c$ power counting rules.}
\end{figure}

The baryon $1/N_c$ expansion has yielded many results for the spin-flavor properties of the ground state baryons, consisting of the spin-$1/2$ $SU(3)$ flavor octet ${\bf 8}$ and the spin-$3/2$ decuplet ${\bf 10}$~\cite{djm1,djm2}, as well as for excited baryons and baryons containing a single heavy quark $Q=c,b$.  In the following section, I discuss one particular example, namely the masses of heavy quark baryons, a subject which continues to be of intense experimental interest. 

\section{$Qqq$ Baryon Masses}

The ground state baryons containing a single heavy quark form a representation of contracted light-quark spin-flavor symmetry.  This representation decomposes into the spin and light-quark flavor $SU(3)$ representations given in Figure~2.  In addition to light-quark spin-flavor
symmetry, heavy-quark baryons satisfy heavy-quark spin-flavor symmetry in the heavy quark limit $m_Q \rightarrow \infty$ and in the large-$N_c$ limit.  Both of these spin-flavor symmetries are broken. 
Corrections to approximate light-quark spin-flavor symmetry are proportional to powers of $SU(3)$ flavor symmetry breaking and powers of $1/N_c$, whereas  
deviations from heavy quark spin-flavor symmetry are suppressed by powers of $\Lambda_{\rm QCD}/m_Q$ and $1/N_c$~\cite{j,h}.

\begin{figure}
{\begin{picture}(0,100)(-100,0)
\put(-75,70){$\Lambda_Q$}
\put(-100,40){$\Xi_Q$}
\put(-50,40){$\Xi_Q$}
\end{picture}}
\hspace{2.cm}
{\begin{picture}(0,100)(-100,0)
\put(-75,70){$[ud]Q$}
\put(-75,40){$[qs]Q$}
\end{picture}}
\hspace{4.cm}
{\begin{picture}(0,100)(-100,0)
\put(-125,70){$\Sigma_Q^{(*)}$}
\put(-75,70){$\Sigma_Q^{(*)}$}
\put(-25,70){$\Sigma_Q^{(*)}$}
\put(-100,40){$\Xi_Q^{\prime(*)}$}
\put(-50,40){$\Xi_Q^{\prime(*)}$}
\put(-75,10){$\Omega_Q^{(*)}$}
\end{picture}}
\hspace{3.cm}
{\begin{picture}(0,100)(-100,0)
\put(-75,70){$\{qq\}Q$}
\put(-75,40){$\{qs\}Q$}
\put(-75,10){$\{ss\}Q$}
\end{picture}}
\caption{Ground state baryons containing a single heavy quark $Q=c,b$ and light quarks $q=u,d$ and $s$.  The spin-flavor baryon representation consists of
a spin-$1/2$ flavor $SU(3)$ antitriplet ${\bf \bar 3}$, containing isospin multiplets $\Lambda_Q$ and $\Xi_Q$; a spin-$1/2$ $\bf 6$, containing
isomultiplets $\Sigma_Q$, $\Xi^\prime_Q$ and $\Omega_Q$; and a spin-$3/2$ $\bf 6$, containing isomultiplets
$\Sigma_Q^*$, $\Xi_Q^*$ and $\Omega_Q^*$.}
\end{figure}
 
   The form of the $1/N_c$ expansion for heavy-quark baryons was derived in Ref.~\cite{h}, and applied to the case of heavy quark baryon masses.  A hierarchy of mass combinations in powers of $1/N_c$, $SU(3)$ flavor breaking, and $\Lambda_{\rm QCD}/m_Q$ was obtained, resulting in a series of baryon mass relations of varying accuracy.
The most suppressed mass relations are predicted to hold at the sub-MeV level, and hence are essentially exact mass
relations.  An example of a highly suppressed mass combination is the flavor-${\bf 27}$ combination
\begin{equation}
\frac 1 4 \left[ \left( \Sigma_Q^* - \Sigma_Q \right) - 2 \left( \Xi_Q^* - \Xi_Q^\prime \right) + \left(\Omega_Q^* - \Omega_Q \right) \right],
\end{equation}
which is currently measured to be $-0.9 \pm 1.2$~MeV in the charm system.  It is predicted to be order $\pm 0.2$~MeV for $Q=c$ and a factor of $m_c/m_b$ smaller for $Q=b$.
Mass relations of several different types also were obtained.  Highly suppressed mass combinations involving bottom and charm baryon masses are found, such as the {\it difference} of
\begin{equation}
\frac 1 6 \left[ \left( \Sigma_Q + 2 \Sigma_Q^* \right)-  2 \left( \Xi_Q^\prime + 2 \Xi_Q^* \right) + \left( \Omega_Q + 2 \Omega_Q^* \right) \right] 
\end{equation}
for $Q=b$ and $Q=c$ baryons, which is predicted to be sub-MeV.  Further relations hold between
heavy-quark mass splittings and mass splittings of baryons which do not contain a heavy quark $Q$, such as
\begin{equation}
\left[ \frac 1 3 \left( \Sigma_Q + 2 \Sigma_Q^* \right) - \Lambda_Q \right] = \frac 2 3 \left( \Delta - N \right) .
\end{equation}
The most suppressed mass combinations have been used to predict unmeasured heavy-quark baryon masses in terms of measured ones. 

\begin{table}
\begin{tabular}{lr}
\hline
  \tablehead{1}{r}{b}{Theory}
  & \tablehead{1}{r}{b}{Experiment (Year)} \\ 
\hline
$\Xi_c^\prime = 2580.8 \pm 2.1\ $ & $\Xi_c^\prime = 2576.5 \pm 2.3\ (1999)$ \\
$\Omega_c^* = 2760.5 \pm 4.9\ $ & $\Omega_c^* = 2768.3 \pm 3.0\ (2006)$ \\
$\Xi_b = 5805.7 \pm 8.1\ $ & 
$\Xi_b^- = 5774 \pm 11 \pm 15\ (2007)$\\ 
& $\Xi_b^- = 5792.9 \pm 2.5 \pm 1.7\ (2007)$ \\
$\Sigma_b = 5824.2 \pm 9.0\ $ & $\Sigma_b = 5811.5 \pm 1.7\ (2007)$ \\
$\Sigma_b^* = 5840.0 \pm 8.8 \ $ & $\Sigma_b^* = 5832.7 \pm 1.8\ (2007)$\\
$\Omega_b = 6039.1 \pm 8.3 \ $ & $\Omega_b^- = 6165 \pm 10 \pm 13 \ (2008)$
\tablenote{$D0$ value~\cite{D0}} 
\\
& $\Omega_b^- = 6054.4 \pm 6.8 \pm 0.9 \ (2009)$
\tablenote{CDF value~\cite{CDF}} 
\\
\hline
\end{tabular}
\caption{Comparison of the $1/N_c$ mass predictions~\cite{h} with experiment.  All masses are in MeV.}
\label{tab:table1}
\end{table}

Table~1 lists the predictions of heavy-quark baryon masses of the $1/N_c$ expansion~\cite{h} together with the subsequently discovered masses and year of discovery.
At the time of the first theoretical work, five of the eight charm baryons and one bottom baryon, $\Lambda_b^0$, had been observed.  Predictions were made for the remaining unobserved $\Sigma_c^*$, $\Xi_c^\prime$ and $\Omega_c^*$ masses.  
In the case of the $\Xi_c^\prime$, the extremely accurate prediction of the $1/N_c$ analysis was later confirmed by experiment.  The predictions for the $\Sigma_c^*$ and $\Omega_c^*$ were correlated, and as soon as the $\Sigma_c^*$ was observed by CLEO, it became possible to predict the $\Omega_c^*$
very accurately.  This prediction was confirmed nine years later by the experimental observation.
It took until summer 2007, over a decade after the theoretical predictions, for additional bottom baryons, $\Xi_b^-$, $\Sigma_b$ and $\Sigma_b^*$, to be discovered at the Tevatron.  All of the measurements are in excellent agreement with the $1/N_c$ predictions.  The situation changed last year when D0 published a discovery paper on the $\Omega_b^-$.  The D0 value was much higher than the theoretical prediction, see Table~1.  Just the week before this conference, however, CDF put out its first paper on the measurement of the 
$\Omega_b^-$ mass, with a value more than $100$~MeV smaller than the D0 value.
The new CDF measurement is in good agreement with the theoretical prediction of the $1/N_c$ expansion.  Finally, predictions have been made in the $1/N_c$ expansion for the remaining unobserved  
$\Xi_b^\prime$, $\Xi_b^*$ and $\Omega_b^*$ baryons~\cite{h}.  It will be interesting to see how the experimental
situation develops from here.

\section{CONCLUSIONS}

The $1/N_c$ expansion for baryons has resulted in many successful predictions for QCD baryons,
and has proved to be an important nonperturbative method for studying the spin-flavor properties of baryons
at low energies.  The $1/N_c$ expansion is performed about the large-$N_c$ limit, where baryons display
enhanced spin-flavor symmetry.  I have focused in this write-up on one application of the $1/N_c$ expansion,
the masses of baryons $Qqq$ containing a single heavy quark $Q=c$ or $b$.  The $1/N_c$ suppression factors
were crucial for making very accurate predictions for heavy-quark baryon masses.  Since all of the charm baryon
masses of the ground state spin-flavor representation have been observed, the complete pattern of masses in the $1/N_c$, $SU(3)$ flavor symmetry breaking and heavy-quark spin-flavor symmetry breaking expansions can be displayed in the charm system~\cite{h}.
The same hierarchy exists for bottom baryons, and it can be used to narrow the experimental search for the
remaining $\Xi_b^\prime$, $\Xi_b^*$ and $\Omega_b^*$ baryons.





\bibliographystyle{aipproc}   

\bibliography{sample}

\IfFileExists{\jobname.bbl}{}
 {\typeout{}
  \typeout{******************************************}
  \typeout{** Please run "bibtex \jobname" to optain}
  \typeout{** the bibliography and then re-run LaTeX}
  \typeout{** twice to fix the references!}
  \typeout{******************************************}
  \typeout{}
 }


\end{document}